\documentclass[prl,aps,twocolumn,superscriptaddress,notitlepage]{revtex4-1}
\usepackage{amssymb,amsfonts,amsmath,amstext,graphicx,hyperref,lipsum,empheq}
\usepackage{float}
\usepackage[normalem]{ulem}
\hypersetup{
	colorlinks=true,
	linkcolor=blue,
	filecolor=magenta,      
	urlcolor=cyan,
}
\usepackage{amsthm}
\theoremstyle{plain}

\newtheorem*{theorem*}{Theorem}

\usepackage[dvipsnames]{xcolor}

\def\lket#1{\vert#1\rangle\hspace{-1mm}\rangle}
\def\lbra#1{\langle\hspace{-1mm}\langle#1\vert}

\def\ket#1{|{#1}\rangle}

\def\BraVert{\egroup\,\mid\,\bgroup}

\definecolor{myblue}{rgb}{.8, .8, 1}

\begin{document}

\title{Role of Coherence and Degeneracies in Quantum Synchronisation} 

\author{Parvinder Solanki}
\email{psolanki@iitb.ac.in}
\affiliation{Department of Physics, Indian Institute of Technology-Bombay, Powai, Mumbai 400076, India.}

\author{Noufal Jaseem}
\affiliation{Department of Physics, Indian Institute of Technology-Bombay, Powai, Mumbai 400076, India.}

\author{Michal Hajdu\v{s}ek}
\affiliation{Keio University Shonan Fujisawa Campus, 5322 Endo, Fujisawa, Kanagawa 252-0882, Japan.}

\author{Sai Vinjanampathy}
\affiliation{Department of Physics, Indian Institute of Technology-Bombay, Powai, Mumbai 400076, India.}
\affiliation{Centre for Quantum Technologies, National University of Singapore, 3 Science Drive 2, 117543 Singapore, Singapore.}
\email{sai@phy.iitb.ac.in}

\begin{abstract}
    Progress on the study of synchronisation in quantum systems has been largely driven by specific examples which resulted in several examples of frequency entrainment as well as mutual synchronisation. Here we study quantum synchronisation by utilising Liouville space perturbation theory. We begin by clarifying the role of centers, symmetries and oscillating coherences in the context of quantum synchronisation.
    We then analyse the eigenspectrum of the Liouville superoperator generating the dynamics of the quantum system and determine the conditions under which synchronisation arises. We apply our framework to derive a powerful relationship between energy conservation, degeneracies and synchronisation in quantum systems. Finally, we demonstrate our approach by analysing two mutually coupled thermal machines and the close relationship between synchronisation and thermodynamic quantities.
\end{abstract}

\maketitle

\date{\today}

\paragraph{Introduction.---}
In analogy to classical synchronisation, quantum synchronisation stands for the adjustment of rhythms of self-sustained oscillators under the effect of weak coupling or an external drive \cite{pikovsky2003synchronization}. Given the ubiquity of synchronisation in the classical world \cite{pikovsky2003synchronization,strogatz2004sync}, quantum synchronisation has garnered increasing interest. Initially, systems whose mean-field dynamics resembled classical dynamical systems were the primary focus of investigation \cite{amitai2017synchronization,manzano2013synchronization,heinrich2011collective,bastidas2015quantum,witthaut2017classical,marquardt2006dynamical,lee2013quantum,walter2014quantum,sonar2018squeezing,lorch2016genuine,nigg2018observing,kwek2018no,kato2019semiclassical,kato2020enhancement}. Following this, the notion of phase-locking was soon extended to genuinely quantum systems with no classical counterparts \cite{roulet2018synchronizing,roulet2018quantum,noufal2020pre}. The primary concern of the early studies of quantum synchronisation was to motivate a system-specific measure of synchronisation and then demonstrate that such measure attains a finite value in some region of the parameter space, effectively demonstrating that synchronisation is possible even in the quantum regime. A typical first step to study synchronization in both classical and quantum systems involves establishing a valid limit cycle dynamics in the absence of perturbation (driving/coupling). Since perturbation of the system can deform its limit cycle and furthermore induce phase locking, measures of synchronisation need to account for this \cite{noufal2020prr}.

What has been lacking thus far is a systematic study of the specific structure of perturbative driving or coupling that could bring about quantum synchronisation. In classical and hence in quantum dynamics, there are slightly different but related definitions of synchronisation \cite{REN20083195, strogatz2018nonlinear,pikovsky2003synchronization,chao1994}. In this manuscript, we begin by clarifying various definitions of synchronisation in quantum systems and construct a systematic perturbation theory in Liouville space to understand phase space based measures of synchronisation. We demonstrate that synchronisation, degeneracies and energy-conserving interactions are intimately linked. We prove a powerful relationship between the energy structure of the physical systems and their interactions that may prevent the systems from synchronising. Finally, we apply our framework to the example of two coupled thermal machines to illustrate the role of degeneracy in synchronisation and the performance of quantum thermal machines.

We begin the discussion with the dynamics of open quantum systems given by the Lindblad master equation
\begin{align}\label{eq:lindy}
    \dot{\rho}=-\mathrm{i}[H_0,\rho] +\sum_k \mathcal{D} [\mathcal{O}_k]\rho,
\end{align}
where $H_0$ is the bare Hamiltonian of the system and $\mathcal{D}[\mathcal{O}_k]\rho=\mathcal{O}_k\rho\mathcal{O}_k^{\dagger}-\frac{1}{2}\{\mathcal{O}_k^{\dagger}\mathcal{O}_k,\rho\}$ models the baths attached to the system.
The steady state $\rho_{ss}$ of the evolution in Eq.~(\ref{eq:lindy}) satisfies $\dot{\rho}_{ss} = 0$.
We focus on the most natural form where the limit cycle is diagonal in the eigenbasis of the bare Hamiltonian, though more general definitions of limit-cycles can be accommodated \cite{noufal2020prr}.
The lack of off-diagonal terms in this basis signifies the absence of any phase-locking, either to an external drive or other systems. Diagonal limit cycles arise naturally in systems coupled to thermal baths \cite{noufal2020pre}. Perturbative driving or coupling to other systems may give rise to synchronisation as measured by a number of various phase-space based measures \cite{roulet2018quantum,roulet2018synchronizing,ameri2015mutual,walter2014quantum,walter2015quantum}.
A common feature of all such measures is that they detect synchronisation for steady states which contain finite off-diagonal terms, pointing at a relationship between coherence generation and synchronisation. This relationship was solidified by the fact that the relative entropy of coherence $\mathcal{S}_{coh}$ \cite{baumgratz2014quantifying}, is a suitable measure of synchronisation \cite{noufal2020prr}.

Classical dynamics is strongly affected by constraints placed on Hamiltonian or on dissipative evolution. For instance, the constraint that phase space volume is conserved implies that Hamiltonian systems cannot have attractors. In the quantum context, a natural question that arises is how constraints placed on Hamiltonian or Lindbladian evolution affect quantum synchronisation. The question is important because quantum information theoretic tasks often are accompanied by a fixed resource constraint, which then affects the underlying dynamics in a non-trivial manner. An example of such a resource constraint is the demand that the coupling between thermal machines should be energy conserving. In this manuscript, we show that degeneracy of the bare Hamiltonian $H_0$ plays a crucial role in understanding the relationship between synchronisation and energy conservation and demonstrate that non-degenerate systems cannot be both energy-conserving and display synchronisation at the same time.

\paragraph{Liouville Eigenspectrum \& Synchronisation.---}
Consider the Lindblad master equation in Liouville form namely
\begin{eqnarray}
\lket{\dot{\rho}(t)}=\mathcal{L}\lket{\rho(t)},
\end{eqnarray}
where $\mathcal{L}$ is the Liouville superoperator and $\lket{\rho(t)}$ is the corresponding Liouville state \cite{suri2018speeding,manzano2020short,gyamfi2020fundamentals}. Eigenvalues of Liouville superoperators are complex numbers $\lambda_\mu=\alpha_\mu+i\beta_\mu$ with physical density matrices corresponding to $\alpha_\mu\leq 0$. The steady state of this master equation is given by the converging solution of the equation above and are given by $\alpha_\mu=0$, with non-zero $\beta_\mu$ indicating oscillating coherences. Before we discuss synchronisation in the context of the Liouvillian eigenspectrum, we note that a prerequisite for phase-space based measure of synchronisation is the existence of a self-sustained oscillator with a valid limit-cycle corresponding to an observable free phase. This disqualifies linear quantum systems and also precludes a discussion of qubit synchronisation \cite{roulet2018quantum}. In classical mechanics, a stable limit-cycle is first established for a non-linear system possessing free phase with neutral stability, which is then synchronised even by a perturbative drive or coupling \cite{pikovsky2003synchronization}. Following these arguments, the limit-cycle in the quantum context is established by proving that the unperturbed Liouvillian $\mathcal{L}_0$ has a diagonal steady state \cite{roulet2018quantum,noufal2020pre,noufal2020prr}. Once the limit cycle is established, the system is typically perturbed by an external drive or coupling to another system, which we represent by $\epsilon \mathcal{L}_V$. Such a perturbation can either produce steady state coherence ($\lambda_k=0$ with the corresponding eigenket having non-zero coherences) or oscillating coherences ($\lambda_\mu=\pm i\beta_\mu$). As discussed below, we show that while steady state coherence can be used to define a phase space based measure of synchronisation, the case of oscillating coherences is more subtle.

\begin{figure}[!htp]
    \includegraphics[width=0.5\textwidth,keepaspectratio]{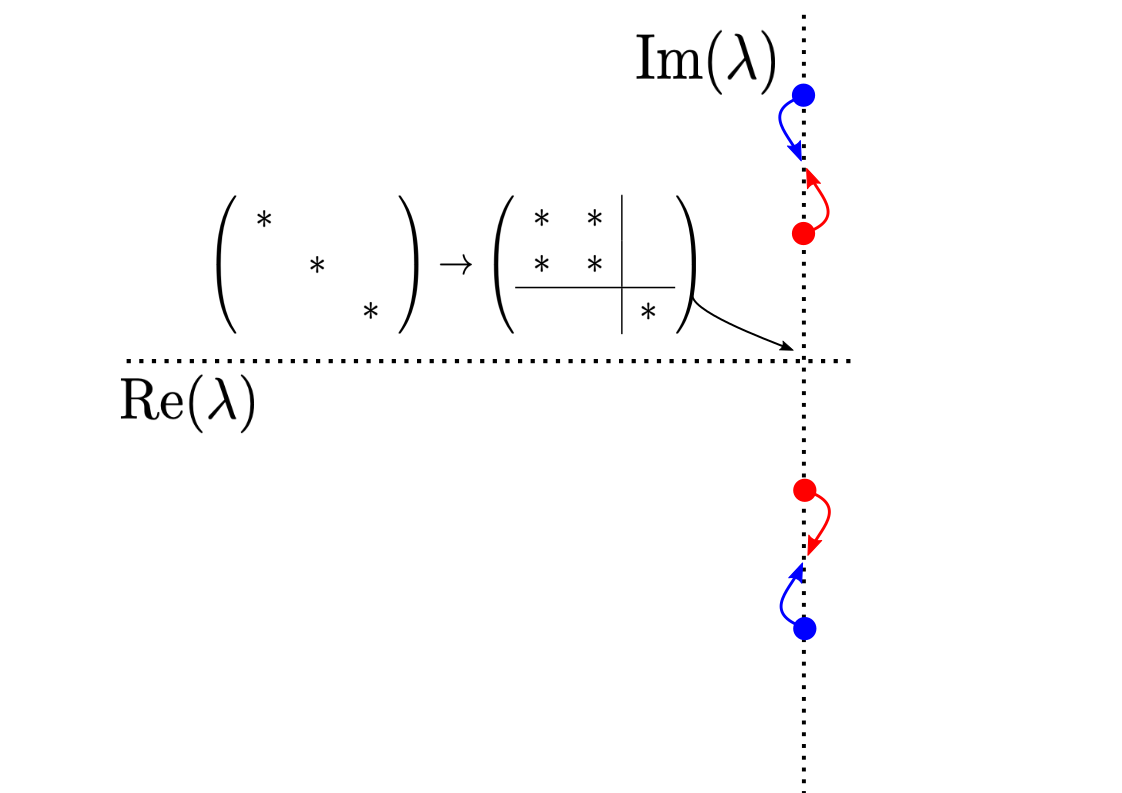}
    \caption{Two different dynamical phenomena that can cause adjustment of rhythms: The first phenomenon is phase synchronisation caused by the development of steady state coherences due to driving or coupling and is indicated by the matrix whose corresponding eigenvalue is zero. The second is a phase ordering phenomenon which we refer to as coherence synchronisation, caused by the adjustment of oscillating coherences (blue and red) indicated by the arrows pulling the imaginary eigenvalues closer to each other.}
    \label{fig:figure1}
\end{figure}

Alternative to the phase space-based measures, one could start with two or more systems whose uncoupled dynamics have steady state oscillating coherences and demand that the oscillations determined by the imaginary eigenvalues adjust their rhythms due to coupling \cite{tindall2020quantum}. This phenomenon, shown in Fig.~\ref{fig:figure1} restricts the underlying limit cycles to themselves be oscillating coherences \cite{btc,anna2020time} and hence can include certain qubit models. We note that the classical analogue of oscillating coherences is centers in classical two-dimensional flows \cite{strogatz2018nonlinear}, which are also complex conjugate pairs of imaginary eigenvalues. Since centers do not represent either a stable limit cycle or a neutral free phase, we do not consider quantum analogues of such systems in this manuscript.

Another issue that arises in defining synchronisation around centers relates to the information preserving nature of the dynamics. Genuine limit cycle dynamics involves a basin of attraction, wherein any dynamics that starts in the basin ends up on the limit cycle. Thus classical limit cycle dynamics does not preserve initial conditions and flows to a unique asymptotic state. This is not the case for quantum dynamics which lead to oscillating coherences, which can arise with strong or weak symmetries \cite{buvca2012note} or when the Lindbladian pumps the quantum system to a subspace with Hamiltonian dynamics \cite{thingna2021degenerated}. For Lindblad dynamics with strong symmetries (i.e., existence of unitary $S$ such that $[S,H]=0$ and $[S,\mathcal{O}_k]=0$), the dimensionality of the Liouville subspace $\mathbb{L}_{ss}$ is determined by the number of unique eigenvalues of $S$ and subspaces with $\mathbb{L}_{ss}\geq2$ are known to be \textit{information preserving} \cite{albert2014symmetries}. For all of these reasons we refer to the phenomenon involving multiple oscillating coherences as undergoing \textit{coherence synchronisation} in general \cite{buca2021algebraic} and use the phrase \textit{phase synchronisation} for models where the underlying dynamics exhibits a stable limit cycle and a neutral free phase. We note that the general theory of phase ordering of oscillating coherences remains an open problem.

Finally we note that the presence of degeneracies has a profound effect on the entire dynamics that has to be carefully reconciled. Lindblad-type master equations are derived under the BMS approximation \cite{breuer2002theory} which may not hold in the presence of degeneracies and can hence modify the underlying master equation dramatically \cite{thingna2014improved,mccauley2020accurate}.

\paragraph{Perturbation of limit cycle oscillators.---}
Having clarified the definition of limit-cycle oscillators in the quantum context, we are now in a position to construct a formal perturbation theory to study phase synchronisation. After adding interaction Hamiltonian $\varepsilon V$ as  a perturbation to $H_0$, total Liouville super-operator can be written as $\mathcal{L}=\mathcal{L}_0+\varepsilon \mathcal{L}_V$ where  $\mathcal{L}_0=\mathcal{L}_{H_0}+\mathcal{L}_D$ is the unperturbed Liouvillian super-operator and $\mathcal{L}_V$ corresponds to $V$.
In general $\mathcal{L}$ is not Hermitian ($\mathcal{L}^{\dagger}\neq \mathcal{L}$) and has different left ($\lket{l_{\mu}}$) and right ($\lket{r_{\mu}}$) eigenvectors \cite{li2014perturbative} such that
\begin{eqnarray}
\mathcal{L} \lket{r_{\mu}}=\lambda_{\mu} \lket{r_{\mu}}, \qquad
\mathcal{L}^{\dagger} \lket{l_{\mu}}=(\lambda_{\mu})^{*} \lket{l_{\mu}}. \label{eq:left_right}
\end{eqnarray}
These eigenvectors do not form an orthonormal basis.
For small $\varepsilon$ we can expand the eigenvectors ($\lket{r_{\mu}}, \lket{l_{\mu}}$) and eigenvalues ($\lambda_{\mu}$) in term of the unperturbed eigenvectors ($\lket{r_{\mu}^{(0)}},\lket{l_{\mu}^{(0)}}$) and eigenvalues ($\lambda_{\mu}^{(0)}$) as follows:
\begin{eqnarray}
\lambda_{\mu}&=&\lambda_{\mu}^{(0)}+\varepsilon\lambda_{\mu}^{(1)}+\varepsilon^2 \lambda_{\mu}^{(2)}+\ldots \\
\lket{r_{\mu}}&=&\lket{r_{\mu}^{(0)}}+\varepsilon\lket{r_{\mu}^{(1)}}+\varepsilon^2 \lket{r_{\mu}^{(2)}}+\ldots \\
\lket{l_{\mu}}&=&\lket{l_{\mu}^{(0)}}+\varepsilon\lket{l_{\mu}^{(1)}}+\varepsilon^2 \lket{l_{\mu}^{(2)}}+\ldots  .
\end{eqnarray}
Using the above set of equations in Eq.~(\ref{eq:left_right}) and equating first order terms in $\varepsilon$ we get
\begin{equation}
    \mathcal{L}_0\lket{r_{\mu}^{(1)}}+\mathcal{L}_V\lket{r_{\mu}^{(0)}}=\lambda_{\mu}^{(0)}\lket{r_{\mu}^{(1)}}+\lambda_{\mu}^{(1)}\lket{r_{\mu}^{(0)}}.\label{eq:1st_order}
\end{equation}

The perturbation $\varepsilon \mathcal{L}_V$ can have three distinct types of effects on the unperturbed steady state which we discuss for completeness. The first effect of perturbation is that it can change the eigenvectors keeping the corresponding eigenvalues same \cite{li2014perturbative}. Let us assume that first order correction to a eigenvalue $\lambda_{\mu}^{(0)}$ is zero i.e. $\lambda_{\mu}^{(1)}=0$ then Eq.~(\ref{eq:1st_order}) reduces to the following form
\begin{equation}
 \mathcal{L}_0\lket{r_{\mu}^{(1)}}+\mathcal{L}_V\lket{r_{\mu}^{(0)}}=\lambda_{\mu}^{(0)}\lket{r_{\mu}^{(1)}}  ,
\end{equation}
and the first order correction to corresponding eigenvector can be written as
\begin{equation}
    \lket{r_{\mu}^{(1)}}=-(\mathcal{L}_0-\lambda_{\mu}^{(0)})^+\mathcal{L}_V\lket{r_{\mu}^{(0)}},
\end{equation}
where $\mathcal{O}^{+}$ is the Moore-Penrose pseudo-inverse of operator $\mathcal{O}$. Now the first order perturbation to the steady state (lets $\lket{\rho_{ss}^{(1)}}\equiv \lket{r_{0}^{(1)}}$) having eigenvalue $\lambda_{0}^{(0)}=0$, reduces to
\begin{eqnarray}
\lket{\rho^{(1)}_{ss}} &=& (-\mathcal{L}_0^{+}\mathcal{L}_V) \lket{\rho_{ss}^{(0)}},
\label{eq:1st_order}
\end{eqnarray}
where $\lket{\rho_{ss}^{(0)}}\equiv \lket{r_{0}^{(0)}}$ is the steady state of $\mathcal{L}_0$.
Depending on the form of $\mathcal{L}_V$ new steady state $\lket{\rho_{ss}}=\lket{\rho^{(0)}_{ss}}+\lket{\rho^{(1)}_{ss}}$ can contain coherences and we can observe steady synchronization for such perturbation. 

Another possibility is that perturbation $ \varepsilon \mathcal{L}_V $ can change the eigenvalue keeping the eigenvector unchanged. Since we are interested in steady state synchronisation, $\lambda_{\mu}^{(1)}$ can only take on an imaginary contribution, which we prove to be unphysical. Using $\lket{r_{\mu}^{(1)}}=0$ in Eq.~(\ref{eq:1st_order}) we will get
\begin{equation}
    \mathcal{L}_V\lket{r_{\mu}^{(0)}}=\lambda_{\mu}^{(1)}\lket{r_{\mu}^{(0)}}.\label{eq:1st_eval}
\end{equation}
First order correction to the eigenvalue can be calculated by multiplying Eq.~(\ref{eq:1st_eval}) with $\sum_{\nu}\lbra{l_{\nu}^{(0)}}$ to get
\begin{equation}
    \lambda_{\mu}^{(1)}=\frac{\sum_{\nu}\lbra{l_{\nu}^{(0)}}\mathcal{L}_V\lket{r_{\mu}^{(0)}}}{\sum_{\nu}\lbra{l_{\nu}^{(0)}}\lket{r_{\mu}^{(0)}}}.
\end{equation}
 Since we are interested in quantum synchronisation of coupled non-degenerate subsystems, we take the underlying limit-cycle states to be diagonal steady states. 
When complex eigenvalues arise in the Liouville spectrum, they appear as complex-conjugate pairs to preserve positivity and hence  at least two steady states with eigenvalues $i\beta$ and $-i\beta$ will arise with corresponding Liouville eigenvectors $\lket{\rho^+}$ and $\lket{\rho^-}$.
Note that the steady state is going to be a linear combination of both the states i.e. $\lket{\rho}= e^{i\beta t}\lket{\rho^+}+e^{-i\beta t}\lket{\rho^-}$. For $\rho$ to be a valid density matrix it needs to be Hermitian, implying $(\rho^-)^{\dagger}\equiv\rho^+$ and since $\rho^-$ and $\rho^+$ are diagonal density matrices, $\rho^-\equiv\rho^+$. This cannot be true as one eigenvector cannot have two different eigenvalues. This concludes the proof.
A third possible effect of perturbation can be change in both eigenvalues and eigenvectors simultaneously. It is easy to see that this causes the phase synchronisation measures to also oscillate in time, making the phase-space based methods might not be suitable for discussing the subsequent non-linear dynamics of oscillating coherences.

\paragraph{Energy Conservation \& Degeneracies.---}

Until now we have not placed any constraints on $\mathcal{L}_V$. Typical constraints on coupled quantum systems in a thermodynamic setting is to demand energy conserving interactions so that the coupling is not associated with the work output of the thermal machine. We now prove a theorem relating coherence generation, energy conservation and degeneracy of the system for open quantum system dynamics. We also present the closed system analogue of this theorem in Appendix A for pedagogical clarity. 
\begin{theorem*}
In an open quantum system whose dynamics is described by a Markovian master equation with a non-degenerate energy spacings, energy conserving interactions cannot generate phase synchronisation.
\end{theorem*}

\begin{figure*}[!htp]
    \centering
    \includegraphics[width=\textwidth]{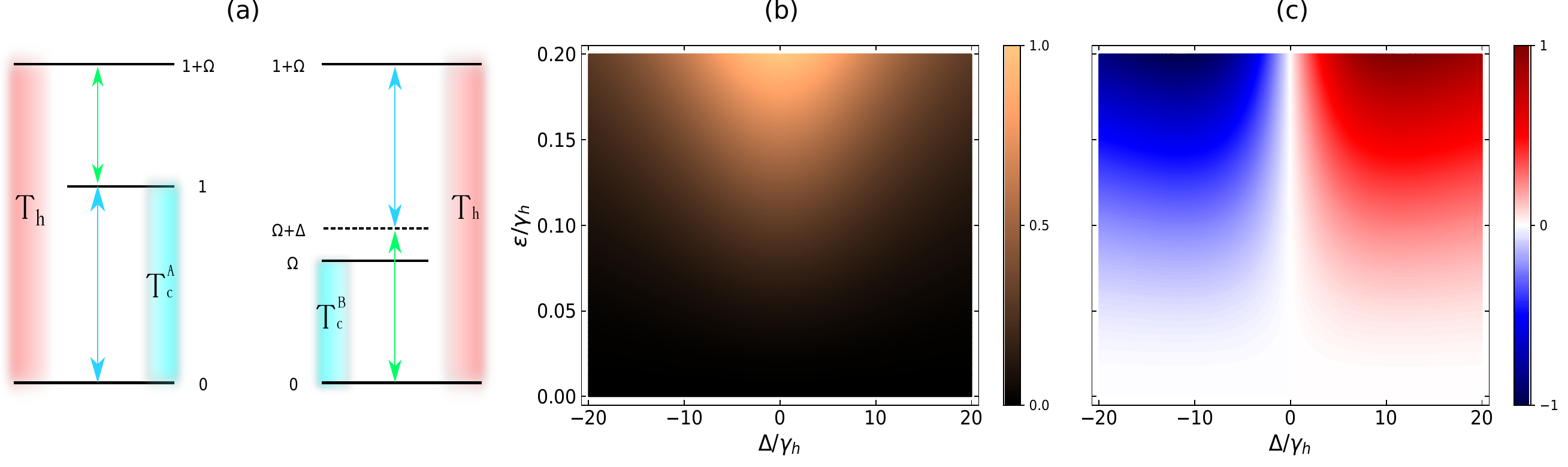}
    \caption{(a) Two coupled heat engines described by Eq.~(\ref{eq:master}) with parameters values given as $\Omega/\gamma_h=40$, $\gamma_h=\gamma_h^{(A,B)}=0.01$,  $\gamma_c=\gamma_c^{(A,B)}=10\gamma_h$,  $\bar{n}_c^{(A,B)}\approx 10^{-3}$ and $\bar{n}_h^{(A,B)}=1$. Plots for steady state (b) sync measure $S_{coh}(\rho)/\text{max}[S_{coh}(\rho)]$ and (c) output power $P/\text{max}[\vert P \vert]$  for a given range of coupling strength $\varepsilon$ and detuning $\Delta$ where $\text{max}[S_{coh}(\rho)]\approx 0.001$ and $\text{max}[\vert P \vert]\approx 1.8\times10^{-6}$.}
    \label{fig:sync_pow}
\end{figure*}

 The proof follows from the fact that coherence generation costs energy in non-degenerate systems \cite{binder2018thermodynamics} and it has been shown that $l_1$ norm of coherence is directly related to synchronisation \cite{noufal2020pre}. This can also be seen structurally using Liouvillian perturbation theory. The first order correction to the unperturbed steady state $\lket{\rho_{ss}^{(0)}}$ of the Lindblad equation is given by Eq.~(\ref{eq:1st_order}). To understand the role of energy conservation we expand $\mathcal{L}_0^{+}$ as a sum of $\mathcal{L}_{H_0}^{+}$ and other terms and rewrite Eq.~(\ref{eq:1st_order}) as
\begin{eqnarray}
\lket{\rho^{(1)}_{ss}} & = & -(\mathcal{L}_{H_0}^{+} + X_{(H_0,\mathcal{D})})\mathcal{L}_V \lket{\rho_{ss}^{(0)}} \nonumber \\
&=& -\mathcal{L}_{H_0}^{+}\mathcal{L}_V \lket{\rho_{ss}^{(0)}} - X_{(H_0,\mathcal{D})}\mathcal{L}_V \lket{\rho_{ss}^{(0)}},
\label{eq:L_1st_order}
\end{eqnarray} 
where $X_{(H_0,\mathcal{D})} \equiv (\delta I-\mathcal{L}^{+}_{H_0}\mathcal{L}_{\mathcal{D}})(\mathcal{L}_{H_0}+\mathcal{L}_{\mathcal{D}})^+$(see Appendix B). This equation forms the basis of our analysis of the relationship between coherence, degeneracies and energy conservation. Since the unperturbed steady state  has the interpretation of the underlying limit cycle state, we demand that $\mathcal{L}_0$ does not generate any coherences. With this background, we now show that energy conserving coupling affects coherence generation. If the bare Hamiltonian is non-degenerate, both $H_0$ and $V$ need to be diagonal to commute. As a consequence of this, the corresponding perturbative corrections to the density matrix are also diagonal in the unperturbed basis, which is already established as a limit cycle state. Consequently no coherence is generated, though the perturbative Hamiltonian can manipulate the underlying limit cycle.

On the other hand, if $H_0$ has degeneracies, the interaction $V$ can be off-diagonal in the degenerate subspace of $H_0$ and still be energy conserving. Hence the corresponding super-operator $\mathcal{L}_V$ will also be off-diagonal only in the degenerate subspace of $\mathcal{L}_{H_0}$ and hence $[\mathcal{L}_{H_0},\mathcal{L}_V]=0$. This means that though $\mathcal{L}_V$ can be off-diagonal, $\mathcal{L}_{H_0}$ and $\mathcal{L}_V$ can share the eigenbasis of $\mathcal{L}_V$.
As $\mathcal{L}_V$ is off-diagonal, $\mathcal{L}_V\lket{\rho_{ss}^{(0)}}\equiv\lket{\tilde{\rho}_{ss}}$ can create coherences.
Since $\mathcal{L}_{H_0}$ is diagonal, the pseudoinverse $\mathcal{L}_{H_0}^{+}=\sum_j \Lambda^{+}_j \lket{\Lambda_j}\lbra{\Lambda_j}$ is also diagonal where $\Lambda^{+}_j=0$ for $\Lambda=0$ and $\Lambda^{+}_j=\Lambda^{-1}_j$ for non-zero $\Lambda_j$.
Therefore the term $\mathcal{L}_{H_0}^{+}\lket{\tilde{\rho_s}}$ can affect coherences only up to a multiplicative factor, whereas $- X_{(H_0,\mathcal{D})}\mathcal{L}_V \lket{\rho_{ss}^{(0)}}$ can generate coherences.
This shows that both terms in Eq.~(\ref{eq:L_1st_order}) are capable of generating coherences. Hence the presence of degeneracies in the bare Hamiltonian can impact coherence generation and synchronisation. It was pointed out earlier that the coherent power output by a (non-degenerate) thermal machine is upper bounded by a measure of synchronisation \cite{noufal2020pre}. We now show how such a relationship is modified in the presence of degeneracies with an example of coupled thermal machines. Several other examples of coupled thermal machines and quantum synchronous systems can be understood within this framework, as presented in Appendix D.

\paragraph{Synchronisation of Coupled Thermal Machines.---}
To exemplify the interplay between degeneracies and coherences, we study the thermodynamic performance and synchronization of two mutually coupled thermal engines shown in Fig.~\ref{fig:sync_pow}(a). We consider two 3-level systems which can be described by bare Hamiltonian $H_0=\sigma_{22}^A+(1+\Omega)\sigma_{33}^A+(\Omega+\Delta)\sigma_{22}^B+(1+\Omega)\sigma_{33}^B$ where $\sigma_{ij}=\vert i \rangle \langle j \vert$, are attached to thermal baths individually.
They are mutually coupled with each other by interaction Hamiltonian of the form $V=\varepsilon(\sigma_{23}^A\sigma_{21}^B+\sigma_{12}^A\sigma_{32}^B+h.c.)$ which is energy conserving interaction for $\Delta = 0 $. The dynamics of such a system is governed by the master equation
\begin{eqnarray}
\dot{\rho}=-i[H_0+V,\rho]+\sum_{i=A,B}(\mathbf{D}_h^i[\rho]+\mathbf{D}_c^i[\rho]),\label{eq:master}
\end{eqnarray}
where $\mathbf{D}^i_{h(c)}[\rho]\equiv \gamma_{h(c)}^i \bar{n}_{h(c)}^i \mathcal{D}[\sigma^{i}_{32}]\rho + \gamma_{h(c)}^i (1+\bar{n}_{h(c)}^i) \mathcal{D}[\sigma^{i}_{23}] \rho$ represents the system $i=(A,B)$ coupled to hot (cold) bath at temperature $T^i_{h(c)}$. The individual systems do not have any degeneracies and the full Hamiltonian has a degeneracy of degree 3 corresponding to the eigenvalue $1+\Omega$. The interaction Hamiltonian $V$ is off-diagonal only in the degenerate subspace, hence $[H_0,V]=0$ and $V$ is an energy conserving coupling. For $\Delta\neq 0$ the degeneracy is lifted and $[H_0,V]\neq 0$, resulting in energy not being conserved anymore. 

Let us study the mutual synchronization for this system. When the thermal engines are not coupled ($\varepsilon=0$) then the  steady states of individual systems under the effects of their thermal bath are diagonal  and hence both systems are in a corresponding limit cycle state. We use the relative entropy of synchronisation \cite{noufal2020prr} $\Omega_R(\rho)=S_{coh}(\rho)=S(\rho_{diag})-S(\rho)$ which measures the distance to the nearest diagonal limit cycle state using relative entropy measure \cite{noufal2020prr}. The synchronization measure $S_{coh}(\rho)$ in Fig.~\ref{fig:sync_pow}(b) displays the typical Arnold tongue behaviour which confirms that steady state mutual synchronization exists between two thermal engines.

 Following definition of power and heat currents \cite{boukobza2006thermodynamics}, power can be given by $P=-i \text{Tr}([H,\rho]H_0)$ where $H=H_0+V$. Hence power for the given system having a steady state $\rho^{ss}=\sum_{i,j=1}^9\rho^{ss}_{ij}\vert i \rangle \langle j \vert $ is given by $P=2g\Delta \text{ Im}[\rho_{35}^{ss}+\rho_{75}^{ss}]$, where $|i\rangle$ denotes the global basis states for the two thermal machines. 
 The heat current from hot (cold) bath is given by $J_{h(c)}=\sum_{i=A,B}\text{Tr}(\mathbf{D}_{h(c)}^i[\rho]H_0)$ (see Appendix C). Power is plotted as a function of coupling strength $\varepsilon$ and detuning parameter $\Delta$ in Fig.~\ref{fig:sync_pow}(c).  For given values of parameters in Fig.~\ref{fig:sync_pow}, $J_h\geq0$ and $J_c\leq0$ where equality holds only for $g=0$. 

Now from Fig.~\ref{fig:sync_pow}, one can observe that for energy conserving case ($\Delta=0$) power output is zero and the synchronization measure is non-zero which means that coherences are generated but no power is consumed or produced for such a case. As soon as we start deviating from $\Delta=0$, energy is no longer conserved and coherences are generated on cost of power being either consumed or produced i.e. $P\neq0$. For $\Delta<0$, power is negative while $J_h\geq 0$ and $J_c\leq 0$ which means that coupled system behaves like a thermal engine and power is generated. It has been shown in \cite{noufal2020pre} that synchronization assists in power generation for thermal engines. Power is positive for $\Delta>0$ which means that power is being consumed by the system while coherence is being generated as a result of energy non-conservation. For $\Delta>0$ the coupled system acts as a dissipator or heater. This example demonstrates our argument that non-degenerate quantum thermal machines cannot be synchronised by energy conserving interactions.

\paragraph{Conclusions.---} 
The literature on quantum synchronisation thus far has been lead by system specific examples. In this manuscript, we go beyond such an approach and discuss a Liouville space perturbation theoretic approach to study phase-space based measures of synchronisation. Unlike previous approaches, we highlight methods to detect emergent synchronicity by analysing the different parts of the Liouville superoperator. While quantum synchronisation of underlying limit cycle oscillators can be understood in terms of steady state coherences under coupling, we also clarified the role of centers, symmetries and oscillating coherences in this context. Finally, we show that degeneracies have a strong role to play in the relationship between thermodynamic quantities such as coherent power and quantum synchronisation. While we assume local master equations without loss of generality due to weak coupling limit, we note that in principle this needs to be reconsidered if the BMS condition no longer applies \cite{local1,local2,local3,local4,local5}. Our example illustrates that while coupled thermal machines can always be synchronised, there is a finite cost to doing so outside of a degenerate manifold. This method can further be used to understand and subsequently design quantum thermal machines and quantum synchronising systems. Our approach can be applied to systematically study quantum synchronisation in the perturbative regime and will find applications in future quantum technologies.

\begin{acknowledgments}
\paragraph{Acknowledgments.---}  SV acknowledges discussions with R. Fazio, P. Parmananda and J. Thingna. SV acknowledges support from a DST-SERB Early Career Research Award (ECR/2018/000957) and DST-QUEST grant number DST/ICPS/QuST/Theme-4/2019. MH acknowledges support by the Air Force Office of Scientific Research under award number FA2386-19-1-4038.
\end{acknowledgments}

	\bibliographystyle{apsrev4-1}
	\color{RoyalBlue}

%

		\color{Black}
\clearpage
\appendix
\section{\\ Appendix }

\subsection{A. Analysis for closed systems}\label{Closed_Proof}
We provide the closed-system analogue of our theorem for pedagogical reasons. For non-degenerate systems, unitary transformations have an energetic cost. Such a cost is captured by the notion of ergotropy \cite{allahverdyan2004maximal}, which is defined as the maximum amount of work that can be extracted from a state $ \rho $ by a unitary transformation $U$. It is defined as $ \mathcal{W}(\rho , H)=\text{Tr}(\rho H)-\text{Tr}(U\rho U^{\dagger} H) $ where $H$ is the Hamiltonian of the system and $U$ is the unitary that maximizes the work extracted. Given that diagonal unitaries like permutations can change the amount of energy stored in a quantum system without inducing coherences, ergotropy has to be divided into two contributions, the latter of which emerges solely due to coherence \cite{francica2020initial}. Since ergotropy is a two-point function, ergotropy-conserving transformations are not energy conserving, though energy conservation always implies ergotropy conservation. To see this, note that we can take a diagonal non-passive state and apply a unitary transformation that final state can have coherences for a fixed ergotropy. This can be done via a two step process, first apply a unitary $U_{pop}$ to make the state passive via permutation of diagonal population and extract out the ergotropy. Use the same amount of energy to apply a second unitary $U_{coh}$ to create coherences in passive state. The total unitary $U_{tot}=U_{coh}U_{pop}$ can generate coherences starting from a diagonal non-passive state with no change in ergotropy. Such uniteries clearly do not imply the energy conservation throughout the process, since $[H,U_{tot}]\neq 0$, although the initial and final states have the same energy. This clarifies the remark about ergotropy conservation not being energy conservation.

For a continuous time evolution of a closed system, energy conservation throughout the process requires $[H(t),U]=0~\forall t$ \cite{NoCost}, where $H(t)=H_0+V(t)$ and $V(t)$ is the interaction Hamiltonian. The unitary evolution from time $t_1$ to $t_2$ can be given by a unitary $U(t)$ of the form
\begin{eqnarray}
U(t)=\mathcal{T}\exp(-i\int^{t2}_{t1}[H_0+V(t)]dt),
\end{eqnarray}
where $\mathcal{T}$ is the time-ordering operator.
So energy conservation also implies $[H_0,V(t)]=0$. If we include degeneracies in the system then the problem becomes more intricate. We employ Brillouin-Wigner perturbation theory (BWPT) \cite{hubavc2010brillouin} to study the form of interaction Hamiltonian $V$ which can generate the coherence under energy conservation paradigm. 

Let us consider a closed system having initial state $\rho_0$ which is diagonal in the basis of the bare Hamiltonian $H_0=\sum_n E^{(0)}_n \vert \lambda^{(0)}_n \rangle \langle \lambda^{(0)}_n \vert$ and is given by
\begin{eqnarray}
\rho_0 \rightarrow \sum_n p^{(0)}_n \vert \lambda^{(0)}_n \rangle \langle \lambda^{(0)}_n \vert \label{rho_0}.
\end{eqnarray}
 Now we turn on a weak interaction $V$ between two subsystems such that total Hamiltonian is given by $H=H_0+\varepsilon V$. Let $\rho$ describe the state of the system after turning on the interaction and given by its spectral decomposition as
\begin{eqnarray}
\rho \rightarrow \sum_n p_n \vert \lambda_n \rangle \langle \lambda_n \vert \label{rho},
\end{eqnarray}
where $\vert \lambda_n \rangle$ are the eigenvectors from the spectral decomposition of $\rho$. As interaction strength is very small then it's effect on the thermal state $\rho_0$ can be studied via perturbation theory. Using BWPT we can write the new eigenvalues of $\rho$ in terms of unperturbed eigenvalues and eigenvectors of $\rho_0$ as
\begin{eqnarray}
p_n = p_n^{(0)} +\varepsilon {p^{(1)}_n} + \varepsilon^2  p^{(2)}_n + \ldots , \label{energy} 
\end{eqnarray} 
such that
\begin{eqnarray}
p_n^{(1)}&=&\langle \lambda_n^{(0)} \vert V \vert \lambda_n^{(0)} \rangle, \nonumber \\
p_n^{(2)}&=&\sum_{m\neq n} \langle \lambda_n^{(0)} \vert V \vert \lambda_m^{(0)} \rangle \frac{1}{p_n-p^{(0)}_m} \langle \lambda_m^{(0)} \vert V \vert \lambda_n^{(0)} \rangle \ldots ,
\end{eqnarray}
where $p_n^{(i=0,1,2,\ldots)}$ are the $ i^{th} $-order corrections in eigenvalues. Similarly from BWPT, eigenvectors of $\rho$ can be expressed in terms of unperturbed eigenvalues and eigenvectors of $\rho_0$ as follow:
\begin{eqnarray}
\vert \lambda_n \rangle &=& \vert \lambda^{(0)}_n \rangle + \varepsilon
 \vert \lambda^{(1)}_n \rangle + \varepsilon^2 \vert \lambda^{(2)}_n \rangle +\ldots,\label{wavefun} 
\end{eqnarray} 
such that
\begin{eqnarray}
\vert \lambda^{(1)}_n \rangle &=& \sum_{m \neq n} \frac{\langle \lambda_m^{(0)} \vert V \vert \lambda_n^{(0)} \rangle}{p_n-p^{(0)}_m}\vert \lambda_m^{(0)} \rangle, \nonumber \\
\vert \lambda^{(2)}_n \rangle &=& \sum_{m\neq n , j \neq n} \frac{\langle \lambda_j^{(0)} \vert V \vert \lambda_m^{(0)} \rangle}{p_n-p^{(0)}_j} \frac{\langle \lambda_m^{(0)} \vert V \vert \lambda_n^{(0)} \rangle}{p_n-p^{(0)}_m} \vert \lambda_j^{(0)}\rangle \ldots,
\end{eqnarray}
where $\vert \lambda^{(i=0,1,2,\ldots)}_n \rangle $ are the $ i^{th} $ order corrections in eigenvectors. Substituting Eq.~(\ref{energy}) and Eq.~(\ref{wavefun}) in Eq.~(\ref{rho}), $\rho$ can be rewritten as
\begin{equation}
\rho = \rho^{(0)} + \varepsilon \rho^{(1)} + \varepsilon^2 \rho^{(2)} \ldots \label{rho_per}
\end{equation} 
where
\begin{eqnarray}
\rho^{(0)}&=&\rho_0=\sum_n p^{(0)}_n \vert \lambda_n^{(0)}\rangle \langle \lambda^{(0)}_n \vert \nonumber  ,\\
\rho^{(1)}&=&\sum_n(p_n^{(1)} \vert \lambda_n^{(0)}\rangle \langle \lambda^{(0)}_n \vert + p_n^{(0)} \vert \lambda_n^{(1)}\rangle \langle \lambda^{(0)}_n \vert \nonumber \\& &+ p_n^{(0)} \vert \lambda_n^{(0)}\rangle \langle \lambda^{(1)}_n \vert), \quad \ldots 
\end{eqnarray}
From Eq.~(\ref{rho_per}) we can see that density matrix $\rho$ is a sum of unperturbed density matrix $\rho_0$ and higher order correction in terms of eigenvector and eigenvalues of $\rho_0$.  Making use of Eq.~(\ref{rho_per}), we will study the generation of coherences in energy conserving and non-conserving interactions.   
\bigskip

If we insist that the energy of a system is conserved at all times, this implies the condition $[H_0,V]=0$. If $H_0$ is non-degenerate then $V$ needs to be diagonal in the eigenbasis of $H_0$ i.e., $V=\sum_n q_n^{(0)}\vert \lambda_n^{(0)}\rangle \langle \lambda_n^{(0)} \vert$. Hence all higher order correction for $\vert \lambda_n \rangle$ are zero because $\langle  \lambda_i^{(0)} \vert V \vert \lambda_j^{(0)} \rangle=q_j^{(0)} \delta_{ij}$. Therefore no coherence can be generated as $\vert \lambda_n \rangle=\vert \lambda^{(0)}_n \rangle$. Only eigenvalues change depending on the values of $p_n^{(1)},p_n^{(2)},\ldots$ and steady state density matrix is given as $\rho=\sum_n (p_n^{(0)}+p_n^{(1)}+p_n^{(2)}+\ldots)\vert \lambda^{(0)}_n \rangle \langle \lambda^{(0)}_n \vert$ which again corresponds to an incoherent state.
\bigskip

On the other hand, if $H_0$ is diagonal and degenerate, then $V$ can be off-diagonal only in the degenerate subspace of $H_0$ and can still commute with $H_0$. This follows from the simple fact that $H_0$ is degenerate and hence is proportional to the identity operator in this space.  This means that though $V$ can be off-diagonal, $H_0$ and $V$ can share the eigenbasis of $V$. Hence following from Eq.~(\ref{wavefun}), the higher order correction ($\vert \lambda_n^{(1)} \rangle,\vert \lambda_n^{(2)} \rangle,\vert \lambda_n^{(3)} \rangle,....$) may be non-zero depending on the form of interaction Hamiltonian $V$. Hence coherences can be generated in degenerate subspace of $H_0$ and the steady state will be a coherent state.

Finally, if $[H_0,V]\neq 0$, then energy of system is not conserved. For this case $H_0$ and $V$ do not share the same eigenbasis. So we can choose $V$ such that higher order correction to $\vert \lambda_n \rangle$ in Eq.~(\ref{wavefun}) may be non-zero at any order of perturbation and hence coherences can be generated at the expense of energy. Hence if a system is non-degenerate then coherences cannot be generated along with energy conservation. 

\subsection{B. Derivation of Eq.~(\ref{eq:L_1st_order})}
The pseudoinverse of the sum of two matrices $A$ and $B$ can be used to define a new operator $X_{(A,B)}$, given by the relation
\begin{equation}
    (A+B)^{+} = A^{+}+X_{(A,B)} \label{dec_X}.
\end{equation}
Using the property of pseudoinverse we can write:
\begin{equation}
    (A^{+}+X_{(A,B)})(A+B) = I \label{P_def},
\end{equation}
where $I$ is the identity operator. Using $A^+A=I_A$, Eq.~(\ref{P_def}) can be rewritten as
\begin{equation}
    I = I_A+A^{+}B+X_{(A,B)}(A+B) \label{P_iden}.
\end{equation}
After rearranging  Eq.~(\ref{P_iden}) we get:
\begin{equation}
    X_{(A,B)} = (\delta I-A^{+}B)(A+B)^{+}, \label{XAB}    
\end{equation}
where $\delta I = I-I_A$. Using Eq.~(\ref{XAB}) of the main text we can write $\mathcal{L}_0^{+}=\mathcal{L}_H^{+}+X_{(H_0,\mathcal{D})}$ where $X_{(H_0,\mathcal{D})} \equiv (\delta I-\mathcal{L}^{+}_{H_0}\mathcal{L}_{\mathcal{D}})(\mathcal{L}_{H_0}+\mathcal{L}_{\mathcal{D}})^+$. Hence Eq.~(\ref{eq:1st_order}) can be re-written as follows
\begin{eqnarray}
\lket{\rho^{(1)}_s} & = & -(\mathcal{L}_{H_0}^{+} + X_{(H_0,\mathcal{D})})\mathcal{L}_V \lket{\rho_s^{(0)}} \nonumber \\
&=& -\mathcal{L}_{H_0}^{+}\mathcal{L}_V \lket{\rho_s^{(0)}} - X_{(H_0,\mathcal{D})}\mathcal{L}_V \lket{\rho_s^{(0)}}.
\end{eqnarray}

\subsection{C. Thermodynamics of coupled engines}
Internal energy of a given system is defined as $\text{Tr}(\rho H_0)$. For a time independent bare Hamiltonian $H_0$, change in internal energy with respect to time is given by $\dot{E}=\text{Tr}[\dot{\rho}H_0]$. We use Eq.~(\ref{eq:master}) to get the $\dot{E}$ for coupled thermal machines which is given as
\begin{eqnarray}
\dot{E}&=&-i\text{Tr}([H,\rho]H_0)+\sum_{i=A,B}\text{Tr}(\mathbf{D}_{h}^i[\rho]H_0) \nonumber \\
    & &+\sum_{i=A,B}\text{Tr}(\mathbf{D}_{c}^i[\rho]H_0).\label{eq:int_energy}
\end{eqnarray}
Following the usual definition of power and heat currents \cite{boukobza2006thermodynamics}, first term in Eq~(\ref{eq:int_energy}) corresponds to the total power $P$ of the system while second and third terms correspond to the heat current $J_h$ and $J_c$, flowing from hot and cold baths to the system. When the system attains steady state $\dot{E}=0$ implying $P+J_h+J_c=0$. Let $\rho^{ss}=\sum_{i,j=1}^9\rho^{ss}_{ij}\vert i \rangle \langle j \vert $ be the steady state of the system written in the tensor basis $\ket{1}=\ket{1}_A\otimes\ket{1}_B,\ket{2}=\ket{1}_A\otimes\ket{2}_B,\ldots\ket{9}=\ket{3}_A\otimes\ket{3}_B$. The steady state power and heat currents are given by
\begin{eqnarray}
    P&=&2g\Delta \text{ Im}[\rho_{35}^{ss}+\rho_{75}^{ss}],\\
    J_h&=&\gamma_h^A(1+\Omega)\{ n_h^A(\rho_{77}^{ss}+\rho_{88}^{ss}+\rho_{99}^{ss})\nonumber\\ & &-(1+n_h^A)(\rho_{11}^{ss}+\rho_{22}^{ss}+\rho_{33}^{ss})\}\nonumber\\
    & &+\gamma_h^B(1+\Omega)\{ n_h^B(\rho_{33}^{ss}+\rho_{66}^{ss}+\rho_{99}^{ss})\nonumber\\ & &-(1+n_h^B)(\rho_{11}^{ss}+\rho_{44}^{ss}+\rho_{77}^{ss})\},\\
    J_c&=&\gamma_c^A\{n_c^A(\rho_{77}^{ss}+\rho_{88}^{ss}+\rho_{99}^{ss})\nonumber \\ & &-(1+n_c^A)(\rho_{44}^{ss}+\rho_{55}^{ss}+\rho_{66}^{ss})\} \nonumber \\ & & + (\Omega+\Delta)\gamma_c^B\{n_c^B(\rho_{33}^{ss}+\rho_{66}^{ss}+\rho_{99}^{ss})\nonumber \\ & &-(1+n_c^B)(\rho_{22}^{ss}+\rho_{55}^{ss}+\rho_{88}^{ss})  \}.
\end{eqnarray}

\subsection{D. Other Examples}

Our theorem allows us to understand how quantum thermal machines and quantum synchronous machines have been designed in the literature. For instance, a coupled autonomous thermal machine which outputs coherent power would have to do so by generating coherences in a degenerate subspace. Likewise, there are examples of quantum synchronous system which exhibit synchronisation of a relative phase. In order to demonstrate the application of our theorem, we consider three such examples , the first two of which deal with synchronisation of spin-1 atoms \cite{roulet2018synchronizing,roulet2018quantum} while the third example is an entanglement-generating thermal machine \cite{tavakoli2018heralded}.

Entrainment of a spin-1 atom to an external drive has been considered in \cite{roulet2018synchronizing},
where the bare Hamiltonian in the frame rotating at the frequency of the external drive is $H_0=(\omega_d-\omega_0) S_z$, where $\omega_d$ and $\omega_0$ are the drive and the spin's natural frequencies, respectively. The driving Hamiltonian is $V=\varepsilon S_y$. It is clear that $H_0$ is non-degenerate and the external drive is energy non-conserving as $[H_0, V] \neq 0$. Therefore this system entrains to the external drive as \cite{roulet2018synchronizing} showed.

A different example may be considered in the case of the bare Hamiltonian for two spin-1 atoms in \cite{roulet2018quantum} giveen by $H_0=\omega_A S_z^A + \omega_B S_z^B$. Their coupling is described by $V=i  \frac{\varepsilon}{2}(S_+^AS_-^B-S_+^BS_-^A)$.
If $\omega_A \neq \omega_B$ then $[H_0,V]\neq 0$ and coherences are generated because of non-conservation of energy and subsequently the two atoms mutually synchronise.
When $\omega_A = \omega_B=\omega$ then $[H_0,V]= 0$ and the energy is conserved.
But $H_0$'s eigenvalues $\pm\omega$ have degeneracy of degree 2, and eigenvalue 0 has a degeneracy of degree 3.
The coupling Hamiltonian $V$ is off-diagonal only in the degenerate subspace of $H_0$.

Finally, we consider an example outside synchronisation literature \cite{tavakoli2018heralded} where the authors considered an entanglement-generating thermal machine consisting of two spin-1 atoms. Each atom is separately coupled to two different thermal baths while atoms  also interact with each other. In this case $H_0=H_A+H_B$, where $H_A= \sigma_{11}^A+(1+\varepsilon)\sigma_{22}^A$ and $H_B= \varepsilon \sigma_{11}^B+(1+\varepsilon)\sigma_{22}^B$.
The interaction is given by $V=g_1 (\sigma_{02}^A\otimes \sigma_{20}^B) + g_2 (\sigma_{12}^A\otimes \sigma_{10}^B) + g_3( \sigma_{10}^A\otimes\sigma_{12}^B) + h.c.$ . In this case $[H_0,V]=0$, hence energy is conserved.
Again $H_0$ has degeneracy of degree 3 for eigenvalue $(1+\varepsilon)$. Here we can clearly see that $V$ is off-diagonal only in the degenerate subspace of $H_0$, which results in the production of coherences exhibited as entanglement generation \cite{tavakoli2018heralded}.

\end{document}